\documentclass[floatfix,twocolumn,aps,prl,amsmath,amssymb]{revtex4}
\usepackage[latin1]{inputenc}
\usepackage[dvips]{graphicx}

\usepackage{dcolumn}
\usepackage{bm}
\usepackage{color}
\usepackage{url}
\usepackage[colorlinks=true ,citecolor=blue]{hyperref}
\usepackage{amsmath,amssymb}

\begin{document}

\title{The Superconducting and Pseudogap Phase Diagram of High-Tc Cuprates \\}

\author{E. C. Marino$^1$, Reginaldo O. C. Junior$^{1}$, Lizardo H. C. M. Nunes$^{2}$,Van S\'ergio Alves$^{3}$ } 
\affiliation{$^1$Instituto de F\'\i sica, Universidade Federal do Rio de Janeiro, C.P. 68528, Rio de Janeiro RJ, 21941-972, Brazil \\$^2$Departamento de Ci\^encias Naturais, Universidade Federal de S\~ao Jo\~ao del Rei, 36301-000 S\~ao Jo\~ao del Rei, MG, Brazil \\
$^3$Faculdade de F\'\i sica, Universidade Federal do Par\'a, Av.~Augusto Correa 01, Bel\'em PA, 66075-110, Brazil. }

\date{\today}

\begin{abstract}
We derive analytic expressions for the critical temperatures of the superconducting (SC) and pseudogap (PG) phases of the high-Tc cuprates, which are in excellent agreement with the experimental data for single-layered materials such as LSCO, Bi2201 and Hg1201. 
Our effective Hamiltonian, defined in the oxygen square sub-lattices formed by the alternate hybridization of $p_x$ and $p_y$ orbitals with the $3d$ copper orbitals, provides an unified explanation for the $d_{x^2-y^2}$ symmetry of both the SC and PG order parameters. Attractive and repulsive interactions involve holes of the two different sublattices and can be derived from the spin-fermion model. Optimal doping occurs when the chemical potential vanishes. For $N$-layered cuprates, the growth of the optimal temperature with $N$, as well as the trend of the SC and AF domes to superimpose, can be simply understood. Our results for the optimal SC transition temperature are in excellent agreement with the experiments for $N=2$ materials of the $Bi$ and $Hg$ families. For $N=3$ the agreement is still satisfactory, while for $N>3$, it becomes poor. The explanation for these facts allows us to suggest a method for increasing the critical SC temperature in cuprates.\end{abstract}

\maketitle

{\bf Introduction}

Understanding the mechanism of high-Tc superconductivity in the cuprate materials is, at the same time, one of the most fascinating and challenging problems in physics. Thirty years after the experimental discovery of superconductivity in such materials \cite{bm}, we still have several fundamental phenomenological issues of the high-Tc cuprates that cannot be properly accounted for by an underlying theory, despite the enormous experimental and theoretical efforts applied \cite{htsc1,htsc2,htsc3,htsc4,htsc5}. To mention just a few of these issues, let us recall that so far, the specific analytic expression for the curves representing the critical transition temperature as a function of doping, namely, $T_c(x)$, which form the characteristic SC domes in all high-Tc materials, is not known. Also, a theoretical framework that could provide an accurate analytical expression for $T^*(x)$ is not available.

Concerning multi-layered cuprates,
an explanation is still lacking for the fact that the optimal transition critical temperature increases as a function of the number of adjacent $CuO_2$ planes, up to a point and then stabilizes, as one can observe, for instance, in the, $Bi$, $Hg$ and $Tl$ families of cuprates
\cite{honma,honma1}.  
Furthermore, it is not understood why, in multi-layered cuprates with a number, $N$, of adjacent $CuO_2$ planes in the primitive unit cell, the antiferromagnetic (AF) and superconducting (SC) domes found in the phase diagram of such materials come closer to each other and eventually superimpose, as we increase $N$. 

In this study, we adress all the above issues and provide an explanation thereof. For this, we start from an effective Hamiltonian, defined on the oxygen lattice of the $CuO_2$ planes of the cuprates. A crucial feature of our model is the observation that such lattice breaks down into two inequivalent sublattices, for which the $p_x$ or $p_y$ oxygen orbitals, respectively, overlap with the copper $3d$ orbitals. This Hamiltonian describes the kinematics and dynamics of the holes doped into the oxygen ions. Cooper pairs are formed by combining holes belonging to the two different sublattices of the oxygen square lattice, which contain respectively, $p_x$ and $p_y$ orbitals. This naturally leads to a d-wave SC order parameter, which is favored by the attractive interaction sector. A term describing the repulsion between holes, conversely, favors the onset of a non-vanishing d-wave PG order parameter, which results from exciton (electron-hole pair, each belonging to a different sublattice) condensation. Hence, our model naturally provides a unified explanation for the d-wave charater both of the SC and PG order parameters, the latter leading to the DDW scenario \cite{ddw} proposed to explain the PG phenomena. The phase diagram of the cuprates, hence, derives from the duality between the formation of Cooper pair and (DDW) exciton condensates, both with a d-wave symmetry. The two effective interaction terms contained in our Hamiltonian can be derived from a purely magnetic interaction, namely, the spin-fermion model \cite{ecm1}, which is the inspiration for the present approach. 
The doping mechanism is explicitly taken into account by the introduction of a constraint relating the fermion number to a function of the stoichiometric doping parameter.

Quantum dynamical effects are brought up by
functional integrating out the fermion degrees of freedom. This allows one to obtain an effective action in terms of the superconducting order parameter, $\Delta_0$, the pseudogap order parameter $M_0$, the chemical potential $\mu_0$ and the temperature. Then, minimizing the effective potential, which corresponds, to this action we are able to verify that the occurrence of nonzero $\Delta_0$ and $M_0$ are, in general, mutually excludent, thereby indicating a competition between the PG and SC phases. The exception occurs when $g=g_1$, which is the case when one derives the Hamiltonian (\ref{0}) from the spin-fermion system \cite{ecm1}.

 By taking the limits $\Delta_0 \rightarrow 0$ and  $M_0 \rightarrow 0$, respectively, we capture the threshold for the SC and PG transition and thereby arrive at an analytic expression for the critical SC and PG temperatures as a function of doping, namely $T_c(x)$ and $T^*(x)$. This reproduces the familiar SC domes, as well as the PG lines found in the cuprates and is in excellent agreement with the experimental data for single-layerd materials such as LSCO, Bi2201 and Hg1201.
Our results indicate that the optimal amount of stoichiometric doping, $x_0$, which leads to the maximal $T_c$ occurs when the chemical potential vanishes: $\mu_0(x_0)=0$,  $T_c(x) \leq T_c(x_0)$. 

The increase of the optimal temperature, as well as the tendency of the SC and AF domes to superimpose as we increase the number of adjacent planes in the primitive unit cell of multi-layered cuprates, can be simply understood within our approach, as we show that the effective coupling parameter $g$ is enhanced by the number, $N$, of such planes: $g \rightarrow Ng$. \\
\bigskip

{\bf Methods}\\
\bigskip

{\bf A) The Mechanism of Doping}

\begin{figure}
	[h]
	\centerline
	{
		\includegraphics[scale=0.4]{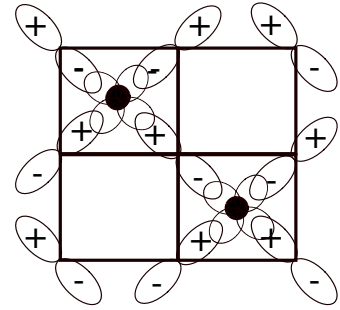}
	}
	\caption{The two sublattices of oxygen ions, formed, respectively, by $p_x$ and $p_y$ orbitals that overlap the $Cu^{++}$ d-orbital. Notice that the $Cu^{++}$-assited overlap of $p_x$ and $p_y$ orbitals along the $x$-direction (of the oxygen lattice) gives a positive sign, whereas the one along the $y$-direction gives a negative sign. Black dots are the $Cu^{++}$ ions.}
	\label{fig01}
\end{figure}

An outstanding feature of all High-Tc cuprates is the presence of one or more $CuO_2$ planes, intercepting the primitive unit cell of such compounds. The $CuO_2$ planes have a lattice structure in which $Cu^{++}$ ions occupy the sites and oxygen ions the links of a square lattice, with a lattice parameter ${\textbf a}'=3.8 $\AA .
These ions are in a $3d^9$ electronic configuration, which results in one spin 1/2 per site. The system of copper ions is a Mott-Hubbard insulator, hence, from this point of view, it forms an array of localized spins
interacting with the nearest-neighbors through the super-exchange mechanism. This structure is ultimately responsible for the antiferromagnetic properties observed in the high-Tc cuprates.  From the point of view of the oxygen ions, however, the picture is different. Indeed,
the oxygen ions are themselves, placed on the sites of a square lattice, with a lattice parameter ${\textbf a}=\sqrt{2} {\textbf a}' / 2 = 1.9\sqrt{2}$ \AA \, which possesses two sublattices, containing, respectively, $p_x$ and $p_y$ oxygen orbitals, which overlap with the $Cu^{++}$ d-orbitals (see Fig.\ref{fig01}), thereby forming bridges that will allow not only hole hopping along the whole oxygen lattice, but also the formation of Cooper pairs as well as excitons along these bridges. As we shall see, this fact naturally explains why both the SC and PG gaps have a d-wave symmetry.

In the case of the pure parent compounds the oxygen ions are doubly charged, namely:
$O^{--}$. Such ions are in a $2p^6$ configuration and the $p_x$ and $p_y$ orbitals contain two electrons each. The valence band which corresponds to the above described oxygen structure contains two electrons per site and, therefore, is completely filled. The electron density is $N_e=\frac{2}{A}$, where $A=a^2$.
As doping is introduced, through some stoichiometric process, parametrized by $x$, one of the two electrons, either from the $p_x$ or the $p_y$ oxygen sublattices is pulled out of the plane, thereby creating a hole in such orbital. Expressing the average hole density per site in the oxygen lattice as $N_h=\frac{2}{A}y$, where $y\in [0,1]$, it follows that the average electron density becomes $N_e=\frac{2}{A}(1-y)$. Now, one must consider that the relation between the stoichiometric doping parameter, $x$ and the average number of holes per site in the oxygen lattice of the $CuO_2$ planes, associated to the $y$-parameter, is not universally known, in general; usually exhibiting different forms for each of the cuprate materials \cite{honma0,honma,honma1}. Furthermore, the presence of interactions should influence such relation. Consequently, we have the hole density parameter, $y$, given by some non-universal function of the doping parameter: $y=f(x)$. We typically do not know the function $f(x)$, therefore, we will describe the doping process through a constraint relating the fermion number directly to the stoichiometric doping parameter $x$, rather then to the density of holes in the oxygen lattice, which is parametrized by
$y$. As we increase the doping parameter $x$, the number of holes in the oxygen lattice will somehow increase as well, eventually reaching an amount where the critical SC temperature reaches a maximum. We call $x_0$ the value of the doping parameter for which this happens. As we will see, the chemical potential will vanish precisely at $x=x_0$.  \\
\bigskip

{\bf B) The Model} 

The underlying pairing mechanism responsible for the superconductivity in cuprates, whatever it may be, must generate an effective hole-attractive interaction Hamiltonian on the oxygen square lattice. A derivation of such a term from a purely magnetic interaction was provided in \cite{ecm1}. There, the hole-attractive interaction comes accompanied by a hole-repulsive term and the competition of both effectively governs the electrons and holes. An attentive analysis must consider the fact that the $Cu^{++}$ ions form bridges between the two oxygen sublattices by overlaping the corresponding $p_x$ and $p_y$ orbitals.
It follows that both the hopping and the interaction of the corresponding oxygen holes (see Fig \ref{fig01}), thereby assisted by the $Cu^{++}$ ions, must involve the two different $p_x$ and $p_y$ oxygen sublattices.
We are going to use a simple Hamiltonian, inspired in \cite{ecm1}, consisting of a hopping term, a hole-attractive and a hole-repulsive terms, that altogether capture these features, namely,
\begin{eqnarray}
&H&=H_0+H_{SC}+H_{PG} 
\nonumber \\
\nonumber \\
&H&=-t \sum_{\textbf{R},\textbf{d}_i} \psi_{B\sigma}^\dagger(\textbf{R}+\textbf{d}_i)\psi_{A\sigma}(\textbf{R})+hc
\nonumber \\
&-& g\sum_{\textbf{R},\textbf{d}_i} \Big [\psi_{B\uparrow}^\dagger(\textbf{R}+\textbf{d}_i)\psi_{A\downarrow}^\dagger(\textbf{R})
 - \psi_{B\downarrow}^\dagger(\textbf{R}+\textbf{d}_i)
\psi^\dagger_{A\uparrow}(\textbf{R})\Big]
\nonumber \\
&\times&
\Big [\psi_{A\downarrow}(\textbf{R})\psi_{B\uparrow}(\textbf{R}+\textbf{d}_i)
 -\psi_{A\uparrow}(\textbf{R}) \psi_{B\downarrow}(\textbf{R}+\textbf{d}_i)\Big]+
\nonumber \\
& g_1&
\sum_{\textbf{R},\textbf{d}_i} \Big [\psi_{B,\sigma}^\dagger(\textbf{R}+\textbf{d}_i)\psi_{A,\sigma'}(\textbf{R})\Big]
 \Big [\psi_{A,\sigma'}^\dagger(\textbf{R})\psi_{B,\sigma}(\textbf{R}+\textbf{d}_i)  \Big] 
\nonumber \\
\label{0}
\end{eqnarray}

In the above expressions, $\textbf{R}$ denotes the sites of a square lattice and $\textbf{d}_i$, $i=1,...,4$, its nearest neighbors. $ \psi_{A,B\sigma}^\dagger$ is the creation operator of a hole, or, equivalently, the destruction operator of an electron, with spin $\sigma=\uparrow,\downarrow$ in sublattice $A,B$. Such sublattices are formed as follows: each oxygen ion possesses one $p_x$ and one $p_y$ orbitals but only one of them hybridizes with the copper 3d orbitals, alternatively, either $p_x$ or $p_y$. Two inequivalent oxygen sublattices are thereby formed, one having hybridized $p_x$ orbitals and the other having $p_y$.
 $t$ is the usual hopping parameter, $g$, the hole-attractive interaction coupling parameter and $g_1$, the hole-repulsive coupling parameter. When this model is derived from the spin-fermion model, one obtains the particular case where $g=g_1$
\cite{ecm1}.

This can be written, up to a constant, in trilinear form, in terms of the Hubbard-Stratonovitch fields $\Phi$ and $\chi$, namely,

\begin{eqnarray}
&H&=-t \sum_{\textbf{R},\textbf{d}_i} \psi_{B\sigma}^\dagger(\textbf{R}+\textbf{d}_i)\psi_{A\sigma}(\textbf{R})+hc
\nonumber \\
&+& \sum_{\textbf{R},\textbf{d}_i}\Phi(\textbf{d}_i) \Big [\psi_{B\uparrow}^\dagger(\textbf{R}+\textbf{d}_i)
\psi^\dagger_{A\downarrow}(\textbf{R}) - \psi_{B\downarrow}^\dagger(\textbf{R}+\textbf{d}_i)
\psi^\dagger_{A\uparrow}(\textbf{R})\Big] 
\nonumber \\
&+& hc
\nonumber \\
&+i& 
\sum_{\textbf{R},\textbf{d}_i}\chi(\textbf{d}_i) \Big [\psi_{B,\sigma}^\dagger(\textbf{R}+\textbf{d}_i)\psi_{A,\sigma'}(\textbf{R}) 
\nonumber \\
&-&
\psi_{A,\sigma'}^\dagger(\textbf{R})\psi_{B,\sigma}(\textbf{R}+\textbf{d}_i)  \Big]
\nonumber \\
&+& 
\frac{1}{g}\sum_{\textbf{R},\textbf{d}_i}\Phi^\dagger(\textbf{R}+\textbf{d}_i) \Phi(\textbf{R}+\textbf{d}_i)
+\frac{1}{2g_1}\sum_{\textbf{R},\textbf{d}_i}\chi^2(\textbf{R}+\textbf{d}_i) ,
\nonumber \\
\label{1a}
\end{eqnarray}

Varying with respect to $\Phi$ and $\chi$, we obtain, respectively,

\begin{eqnarray}
\Phi^\dagger= g\Big [\psi_{B\uparrow}^\dagger
\psi^\dagger_{A\downarrow} - \psi_{B\downarrow}^\dagger
\psi^\dagger_{A\uparrow}\Big] 
\label{1b}
\end{eqnarray}
and
\begin{eqnarray}
\chi = i g_1\Big [\psi_{B,\sigma}^\dagger \psi_{A,\sigma'}- \psi_{A,\sigma'}^\dagger\psi_{B,\sigma}  \Big]
\label{1c}
\end{eqnarray}

$\Phi^\dagger$ is a Cooper pair creation operator, the vacuum expectation value of which, namely, $\Delta=\langle \Phi \rangle$, is a SC order parameter. The PG order parameter, conversely, is $M=\langle \chi \rangle$, where $\chi$ is an exciton creation operator. Cooper pair, as well as exciton formation occurs, respectively, for holes-holes or electron-holes, belonging to different sublattices. 

An XY-asymmetry is naturally in-built, produced by the different signs of overlaping $p_x$ and $p_y$ orbitals along the oxygen lattice x and y directions (see Fig. \ref{fig01}). This imposes the relations $\Delta(\textbf{d}_x)=-\Delta(\textbf{d}_y)$ and $M(\textbf{d}_x)=-M(\textbf{d}_y)$, which, as we shall see, will yield  SC and PG order parameters with a d-wave symmetry.

In momentum space, we have the corresponding Hamiltonian
\begin{eqnarray}
&H&=\sum_{\textbf{k},\sigma}\epsilon(\textbf{k}) \Big [\psi_{B\sigma}^\dagger(\textbf{k})\psi_{A\sigma}(\textbf{k})+hc\Big ]
\nonumber \\
&+& \sum_{\textbf{k}}\Phi(\textbf{k}) \Big [\psi_{B\uparrow}^\dagger(-\textbf{k})
\psi^\dagger_{A\downarrow}(\textbf{k}) - \psi_{B\downarrow}^\dagger(-\textbf{k})
\psi^\dagger_{A\uparrow}(\textbf{k})\Big] +hc
\nonumber \\
&+i&
\sum_{\textbf{k}}\chi(\textbf{k}) \Big [\psi_{A\sigma}^\dagger(-\textbf{k})
\psi_{B\sigma}(\textbf{k}) - \psi_{B\sigma}^\dagger(-\textbf{k})
\psi_{A\sigma}(\textbf{k}) \Big ]
\nonumber \\
&+& \frac{1}{g}\sum_{\textbf{k}}\Phi^\dagger(-\textbf{k}) \Phi(\textbf{k})
+ \frac{1}{2g_1}\sum_{\textbf{k}}\chi(-\textbf{k}) \chi(\textbf{k})
\nonumber \\
\label{1aa}
\end{eqnarray}
where $\epsilon(\textbf{k})=2t [ \cos k_x a + \cos k_y a ]$ is the usual tight-binding energy.


Using the xy-asymmetry of $\Delta(\textbf{d}_i)$, which is caused by the asymmetric overlap of $p_x$ and $p_y$ orbitals described above, we obtain accordingly
\begin{eqnarray} 
\Delta(\textbf{k}) =\sum^4_{\textbf{d}_i=1} \Delta(\textbf{d}_i)
e^{i \textbf{k}\cdot \textbf{d}_i}
=
2\Delta_0 [\cos k_xa-\cos k_ya] 
\label{1c}
\end{eqnarray}
and
\begin{eqnarray} 
M(\textbf{k}) =\sum^4_{\textbf{d}_i=1} M(\textbf{d}_i)
e^{i \textbf{k}\cdot \textbf{d}_i}
=
2M_0 [\cos k_xa-\cos k_ya] ,
\label{1cc}
\end{eqnarray}

both of which have d-wave symmetry. 

Using a mean-field approach for the Cooper pair and exciton fields, $\Phi$ and $\chi$, we obtain the energy eigenvalues $ E(\textbf{k}))=\pm \sqrt{\epsilon^2(\textbf{k})+ |M|^2(\textbf{k}) +|\Delta|^2(\textbf{k})}$. The eigenvalues of $H-\mu\mathcal N$, where $\mu$ is the chemical potential and $\mathcal N$ is the number operator, conversely, are
$ \mathcal{E}(\textbf{k}))=\pm \sqrt{(\sqrt{\epsilon^2(\textbf{k})+| M|^2(\textbf{k})}\pm \mu)^2 +|\Delta|^2(\textbf{k})}$.

\begin{figure}
	[h]
	\centerline
	{
		\includegraphics[scale=0.2]{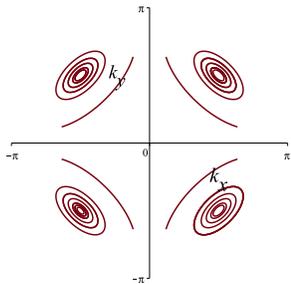}
	}
	\caption{Ellipses or arcs thereof with semi-axes proportional to the total energy are the constant energy curves. For vanishing energy these reduce to the pockets located at the four points $\textbf{K}$.}
	\label{fig02}
\end{figure}

 When we start to dope and, consequently introduce holes in the system, it is natural to expect hole pockets to form around the points where $E^2_{\pm}(\textbf{K}) =0$, namely, $\textbf{K}=(\pm \frac{\pi}{2a},\pm \frac{\pi}{2a})$. By 
making a Taylor expansion of the energy E(\textbf{k})) around the points $\textbf{K}$, it is easy to show that the constant energy curves are the ellipses (or arcs thereof) depicted in Fig. \ref{fig02}. 

This is precisely what is observed in ARPES experiments \cite{arpes}, thus puting our model in a solid experimental basis. 

Notice that all the pseudogap phenomenology, which is explained by the d-wave gap \cite{ddw} including the time-reversal symmetry breakdown and the Nernst effect \cite{ddw1} are accounted for by our model as well.

As we mentioned above, the specific Hamiltonian interaction we use here, can be derived from a spin-fermion system, which describes the multiple magnetic interactions of a system of localized and itinerant spins,  \cite{ecm1,ecm2}. A similar  Hamiltonian is described in \cite{vv}. Our elliptic constant energy curves would coincide with the ones obtained from an asymmetric kinetic Dirac lagrangean \cite{cms}, whereas the corresponding curves obtained from an usual Dirac lagrangean would correspond to circles. All of these must be in the same class of universality, therefore leading to the same phase diagram.

We shall now integrate over the fermions. For this purpose,
we arrange the electron and hole fermion operators of oxygen, with spin $\sigma=\uparrow,\downarrow$ in the form of a four-component Nambu fermion field:

\begin{eqnarray}
\Psi_{a}=\left( \begin{array}{c}
              \psi_{A,\uparrow,a} \\
              \psi_{B,\uparrow,a}  \\
              \psi^\dagger_{A,\downarrow,a} \\
              \psi^\dagger_{B,\downarrow,a}   
\end{array}               \right)
\label{2}
\end{eqnarray}
where the indices, A and B, respectively, denote each of the two $p_x$ and $p_y$ sublattices. 

The cuprates can be classified according to the number of adjacent $CuO_2$ planes contained in their primitive unit cell. The index $a$ indicates to which of the adjacent $CuO_2$ planes the electrons and holes belong. This index runs from $1$ to $N$, where $N=1,2,3,...$, according to the number of planes intersecting the primitive unit cell of the material. In this approach, we shall neglect interplane interactions.
We introduce the $x$-dependence through the constraint
\begin{eqnarray}
\lambda\Big[\sum_{a=1}^N \sum_{C=A,B} \psi^\dagger_{C,\sigma,a} \psi_{C,\sigma,a} - N d(x)\Big ]
\label{3}
\end{eqnarray}
which is enforced by integrating over the Lagrange multiplier field $\lambda$, whose vacuum expectation value is the chemical potential:
 $\langle \lambda \rangle = \mu_0$.
Here $d(x)$ is a function of the stoichiometric doping parameter, to be determined. For consistency we must have $d(0)=\frac{2}{A}$, where $A=a^2$ is the unit cell area of the oxygen lattice: $A=2\ (1.9)^2$ \ \AA$^2$.
 
Conveniently integrating on the complex fields, $\Phi$ and $\chi$, which act as Hubbard-Stratonovitch fields, the Hamiltonian then exhibits two quartic interactions in the fermions, one attractive and another repulsive, respectively with couplings $g$ and $g_1$.
Conversely including the doping constraint and performing the quadratic functional integral over the fermion fields  we obtain the effective Euclidean action $S[\Delta,M,\lambda]$,
\begin{eqnarray}
S[\Delta, M,\lambda] = \int d^2\textbf{r}d\tau \Big [ \frac{\Delta^\dagger \Delta}{g} 
&+&\frac{M^2}{2g_1} +N\lambda d(x) \Big ] +
\nonumber \\
&N& tr \ln \mathcal{A}[\Delta, M,\lambda]
\label{2a}
\end{eqnarray}
where $\mathcal{A}=i\partial_0 - \mathcal{H}+ \lambda  I \otimes (-I)$. (Here $I$ is the $2\times 2$ identity matrix and $H = \Psi^\dagger_a \mathcal{H} \Psi_a$).

 Minimizing this, namely
\begin{eqnarray}
\left[\frac{\delta S}{\delta \Delta}\right]= 
\left[\frac{\delta S}{\delta M}\right]=
\left[\frac{\delta S}{\delta \lambda}\right]_{\Delta=\Delta_0;M=M_0; \lambda=\mu_0}=0
\, ,
\label{7}
\end{eqnarray}
we can determine $\Delta_0,M_0,\mu_0$.

Corresponding to (\ref{7}), we find three equations, namely

\begin{eqnarray}
2\Delta_0\Big [\frac{2T}{\alpha}F(\Delta_0, M_0,\mu_0)-\Big( \frac{\Lambda}{\alpha}-\frac{1}{Ng}\Big)\Big ]=0
\label{7a}
\end{eqnarray}
\begin{eqnarray}
2M_0\Big [\frac{2T}{\alpha}F(\Delta_0, M_0,\mu_0)-\Big( \frac{\Lambda}{\alpha}-\frac{1}{Ng_1}\Big)\Big ]=0
\label{7b}
\end{eqnarray}
and
\begin{eqnarray}
d(x)=\mu_0 \frac{4T}{\alpha}F(\Delta_0, M_0,\mu_0)
\label{7c}
\end{eqnarray}
where $F(\Delta_0, M_0,\mu_0)$ is a function, which, in the regime where $\Delta_0\sim 0 , M_0\sim 0$ is given by
\begin{eqnarray}
&\ &F(\Delta_0 , M_0,\mu_0)\Huge |_{\Delta_0\sim 0 , M_0\sim 0}=\ln2
\nonumber \\
&+&\frac{1}{2}\ln\cosh\Big[\frac{\sqrt{\Delta_0^2+(M_0+\mu_0(x))^2}}{2T}\Big]
\nonumber  \\
&+&\frac{1}{2}\ln\cosh\Big[\frac{\sqrt{\Delta_0^2+(M_0-\mu_0(x))^2}}{2T}\Big]
\nonumber \\
\label{7d}
\end{eqnarray}

In the expressions above, $\alpha=2\pi v_{eff}^2$, and $v_{eff}\simeq 2ta$ is the characteristic velocity and $\Lambda $  is a momentum (energy) characteristic scale, which appears \cite{em} in connection to the characteristic length of the system, namely, the coherence length $\xi$, which essentially measures the range of the pairing interaction (or the Cooper pair size). 
In cuprates we have $\xi\geq \xi_0 \simeq 10$\AA, whereas in conventional superconductors $\xi \geq\xi_0 \simeq 500$\AA.
 The momentum (energy) cutoff is then  $\Lambda  \simeq h v_{eff} /\xi_0=\sqrt{2\pi\alpha}/\xi_0$. It determines the energy scale below which we may consider Cooper pairs as quasiparticles, hence it  must be of the order of $T_c$.

We see that (for $g\neq g_1$) it is, in general, impossible to satisfy (\ref{7a}) and (\ref{7b}) simultaneously with both $\Delta_0\neq 0$ and $M_0 \neq 0$, so we must have either $\Delta_0\neq 0$ and $M_0 = 0$  or $\Delta_0 = 0$ and $M_0 \neq 0$. The first is the SC phase, while the second is the PG phase. The only possibility of having both the SC and PG different from zero would be for $g=g_1$, for which case,
$\Delta_0=M_0\neq 0$ .  \\
\bigskip

{\bf C) The SC Order Parameters and the Critical SC Temperature: $T_c(x)$}

Let us consider firstly the case $\Delta_0\neq 0$ and $M_0 = 0$.

Then (\ref{7a}) and (\ref{7c}) imply 
\begin{eqnarray}
\mu_0(x)=   d(x)\frac{g_c}{2\eta(gN)}  \ \ \ ;\ \ \   \eta(gN)  =\frac{Ng-g_c}{Ng}
\label{11}
		\end{eqnarray}
where $g_c=\alpha/\Lambda$.

In order to find the critical temperature $T_c(x)$, we impose on (\ref{12}) the condition $\Delta_0=0$ and $M_0=0$, which express the fact that the system is in one of the points belonging to the critical curve which separates the SC and PG phases.
Indeed, from (\ref{7a}), we obtain

\begin{eqnarray}
T_c(x)=\lim_{\Delta_0\rightarrow 0}\frac{ \frac{\alpha\eta(gN)}{2g_c}}{F(\Delta_0, M_0=0,\mu_0(x))}
\label{12}.
   \end{eqnarray}

 From (\ref{12}), we see that, for $\Delta_0=0$ and $M_0=0$, the upper bound of $T_c(x)$ occurs at a point $x=x_0$, where $\mu_0(x_0)=0$ and $T_{max}=T_c(x_0)$. Optimal doping occurs when the chemical potential vanishes. According to (\ref{11}), this implies $d(x_0)=0$. The simplest parametrization satisfying this and $d(0)=\frac{2}{A}$ is $d(x)=\frac{2}{Ax_0}(x_0-x)$, such that
\begin{equation}
 \mu_0=2\gamma(gN)(x_0-x),
\end{equation}
with
\begin{equation}
\gamma(gN)=\frac{g_c}{2Ax_0 \eta(gN)}.
\end{equation}
This combined with (\ref{12}) allows us to express the optimal temperature as
\begin{eqnarray}
&\ &T_{max}=\frac{\alpha}{2\ln2}  \frac{\eta(gN)}{g_c} 
\label{14a}
\end{eqnarray}

The critical curve delimiting the boundary of the SC phase, consequently, is obtained from (\ref{12}) and yields
\begin{eqnarray}
\frac{T_c(x)}{\ln2 T_{max}} &=&\lim_{M_0\rightarrow 0} \Big\{\ln2+\frac{1}{2}\ln\cosh\Big[\frac{|M_0+\mu_0(x)|}{2T_c(x)}\Big]
\nonumber  \\
&+&\frac{1}{2}\ln\cosh\Big[\frac{|M_0-\mu_0(x)|}{2T_c(x)}\Big]\Big\}^{-1}
\label{12a}.
   \end{eqnarray}

We see that $T_{max}$ depends linearly on the couplings: $Ng-g_c$, whereas in conventional SC, there is an exponential dependence. This kind of behavior has been extensively studied before \cite{em,ecm2}. 
By using the experimental values of $T_{max}$ for the many different compounds studied here, we find $\Lambda \simeq 0.018 eV$. This is compatible with values of $h v_{eff}$ and $\xi_0$, found in previous studies \cite{emsc}.

Inside the SC phase, we have $M_0=0$. Inserting this condition in (\ref{12}), we can derive an expression for the SC gap as a function of the temperature and doping, which is valid for $T\simeq Tc$  (note that both $\mu_0$ and $T_c$ depend on $x$)
\begin{eqnarray}
\Delta_0^2(T,x)=\Big[2T \cosh^{-1}\Big[\cosh\Big(\frac{\mu_0}{2T_c} \Big) [r]^{\frac{T_c}{T}-1}\Big] \Big]^2-\mu_0^2,
\label{12b}
   \end{eqnarray}
where
\begin{eqnarray}
r = \exp\Big[\ln 2\frac{T_{max}}{T_c(x)} \Big] > 1.
\label{12c}
   \end{eqnarray}
Notice that $\Delta_0(T_c,x)=0$ and, since $r>1$ and $\cosh$ is a monotonically increasing function, we must have
$\Delta_0(T>T_c,x)=0$, while $\Delta_0(T<T_c,x)\neq 0$. \\

\bigskip

{\bf D) The PG Order Parameters and the Critical PG Temperature: $T^*(x)$}

We consider now the case where  $\Delta_0 = 0$ and $M_0 \neq 0$. In order to find the critical temperature $T^*(x)$, we take (\ref{7b}) in the limit $M_0\rightarrow 0$, which leads to
\begin{eqnarray}
T^*(x)=\lim_{M_0\rightarrow 0}\frac{ \frac{\alpha\eta(g_1N)}{2g_c}}{F(\Delta_0=0, M_0,\tilde\mu_0(x))}
\label{12bb}.
   \end{eqnarray}

Now (\ref{7c}) yields the following expression for the chemical potential
\begin{equation}
\tilde\mu_0=2\tilde\gamma(g_1N)(\tilde x_0-x).
\end{equation}
Observe that, because $M_0\neq 0$ in the PG phase, the chemical potential $\tilde\mu_0(x)$, no longer  vanishes at the optimal doping $x_0$. \\



\bigskip

{\bf Results}

We describe now the results obtained by applying our approach firstly to three single-layered cuprates ($N=1$), namely LSCO, Bi2201 and Hg1201, for which, $\gamma\equiv\gamma(gN)$ and $\tilde\gamma\equiv\gamma(g_1N)$, for $N=1$. Thereafter we present  our results concerning multi-layered cuprates. \\
\bigskip
{\bf A) SC Gap}\\
\bigskip
{\bf A.1) LSCO}

Starting from (\ref{12}) we can write (see Suplementary Material)
\begin{eqnarray}
T_c(x) =\frac{\ln2 \ \ T_{max}}{\ln2+\frac{\gamma|x_0-x|}{T_c(x)}+\frac{1}{2}\Big \{ \exp\left[ - \frac{2\gamma|x_0-x|}{T_c(x)} \right] - 1  \Big \}}.
\label{16}
		\end{eqnarray}

The solution $T_c(x)$ of this implicit equation for the critical temperature of the SC transition, obtained with MAPLE, is depicted in Fig. 3.
\begin{figure}[tbh]
	\centering
	\includegraphics[scale=0.30]{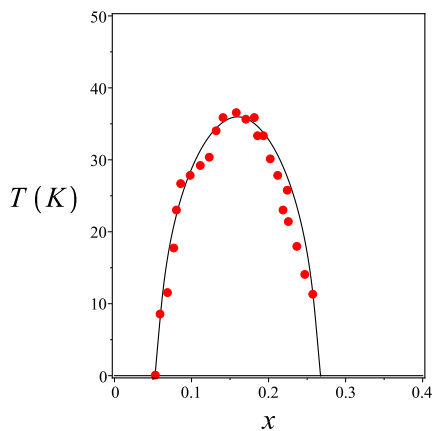}
	\caption{Solution of Eq. (\ref{16}) for the SC dome of LSCO. Experimental data from \cite{1,2,3,4}.} \label{fig1}
\end{figure}

For obtaining this result, we used  $\gamma =0.020\  eV$, $x_0=0.16$ and $T_{max}=0.0031\  eV$ (Notice that$\ T(K)=11604\  T(eV)$).
Observe also, that with the choice of $\gamma$, we adjust only one parameter, namely $g$, of our analytic expression for $T_c$. Entering the experimental values of $T_{max}$ and $x_0$ and using the value of $\Lambda$, we find $g=0.40226 \ eV^{-1}$, $ g_c = 0.30766 \ eV^{-1}$ and $ \eta(1g)=0.23517$.

Taking the limit $T_c \rightarrow 0$ in (\ref{16}), we find the two quantum critical points where the SC dome starts at $T=0$. These are given by
\begin{eqnarray}
x^{\pm}_{SC} = x_0 \pm \frac{T_{max}}{\gamma}\ln 2 = x_0 \pm \alpha A x_0 \frac{\eta^2(gN)}{g^2_c}.
\label{17}
		\end{eqnarray}
Inserting the above numerical values, for $N=1$,  we find: $x^-_{SC} = 0.053$ and $x^+_{SC} = 0.2669$.

\begin{figure}[tbh]
	\centering
	\includegraphics[scale=0.35]{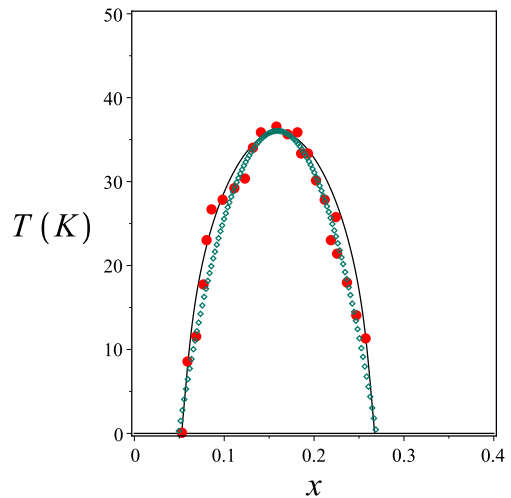}
	\caption{The empirical parabolic fit for the SC dome of LSCO, dotted line, superimposed with our solution for Eq. (\ref{16}), solid line. Experimental data from \cite{1,2,3,4,7}.} 
	\label{fig2b}
\end{figure}

It is instructive to compare our result with the empirical curve, obtained by fitting the data for the LSCO dome, by
the parabola \cite{honma,honma1,emp} 
$$
T_c(x) =T_{max}\left[1- 82.616 (x_0-x)^2\right ].
$$
In Fig. \ref{fig2b} we superimpose it with our solution of
(\ref{16}).
\\
\bigskip

{\bf A.2) Bi2201 and Hg1201}

Now, for Bi2201 and Hg1201, we have different equations for $T_c(x)$ in the underdoped, $x<x_0$ and overdoped,
$x>x_0$ regions. The reason is the order in which we take the limit $M\rightarrow 0$, starting from (\ref{12}), considering that $\mu_0(x)$ has different signs for $x_0 -x>0$ and $x_0 -x < 0$, thus leading to different equations in each region. 
Indeed, we obtain (see Suplementary Material)
\begin{eqnarray}
T_c(x) =\frac{\ln2 \ \ T_{max}}{\ln2+\frac{\gamma|x_0-x|}{T_c(x)}+\frac{1}{2}\Big \{ \exp\left[ - \frac{2\gamma (x_0-x)}{T_c(x)} \right] - 1  \Big \}}
\label{18}
		\end{eqnarray}
for $x<x_0$
and

\begin{eqnarray}
T_c(x) =\frac{\ln2 \ \ T_{max}}{\ln\Big [1+ \exp\left[- \frac{2\gamma (x_0-x)}{T_c(x)} \right]  \Big ]}
\label{19}
		\end{eqnarray}
for $x>x_0$.

\begin{figure}[tbh]
	\centering
	\includegraphics[scale=0.30]{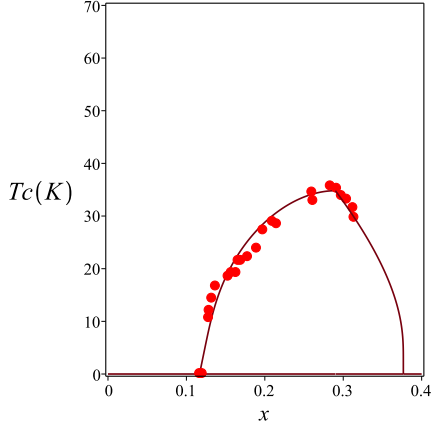}
	\caption{Solution of Eqs. (\ref{18}) and (\ref{19}) for the SC dome of Bi2201. Experimental data from \cite{5,6}.} \label{fig2}
\end{figure}

Let us consider Bi2201 first.
The solution $T_c(x)$ of these implicit equations for the critical temperature of the SC transition of Bi2201, obtained by MAPLE, is depicted in Fig. \ref{fig2}.

\begin{figure}[tbh]
	\centering
	\includegraphics[scale=0.30]{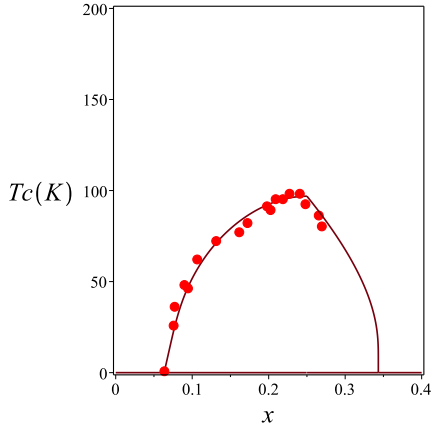}
	\caption{Solution of Eqs. (\ref{18}) and (\ref{19}) for the SC dome of Hg1201. Experimental data from \cite{7,8}.} \label{fig3}
\end{figure}

For obtaining this result, we used $N=1$, $\gamma =0.0120\ eV$, $x_0=0.29$ and $T_{max}=0.0030 \  eV$, which imply $g=0.40800  \ eV^{-1}$ and $g_c = 0.31552  \ eV^{-1}$ and $\eta(N=1)=0.22666 $.

Taking the limit $T_c \rightarrow 0$ in (\ref{16}), we find the two quantum critical points where the SC dome starts at $T=0$. These are given by
\begin{eqnarray}
x^{-}_{SC} = x_0 - \frac{T_{max}}{\gamma}\ln 2\ \ ;\ \ x^{+}_{SC} = x_0 + \frac{T_{max}}{2\gamma}\ln 2.
\label{20}
		\end{eqnarray}
Inserting the above numerical values, we find: $x^-_{SC} = 0.1175$ and $x^+_{SC} = 0.3766$.

Now consider Hg1201.
Then, using Eqs. (\ref{18}) and (\ref{19}) with parameters $\gamma =0.0310\ eV$, $x_0=0.25$ and $T_{max}=0.00835 \ eV$, we obtain
the solution $T_c(x)$ of this implicit equation for the critical temperature of the SC transition, which is depicted in Fig. \ref{fig3}.
The above values imply $ g =2.14781  \ eV^{-1}$ and $ g_c =0.78612   \ eV^{-1}$ and $\eta(1)= 0.63398$.

Inserting the above numerical values in (\ref{20}), we now find: $x^-_{SC} = 0.064$ and $x^+_{SC} = 0.3433$.\\
\bigskip

{\bf B) Pseudogap}\\
\bigskip

We are now going to obtain the upper critical line delimiting the PG phase, namely  $T^*(x)$. For this purpose, we start from (\ref{12bb}) and taking the limit $M_0\rightarrow 0$, obtain
\begin{eqnarray}
T^*(x) =\frac{\frac{\alpha\eta(g_1N)}{2g_c}}{\ln\Big [1+ \exp\left[- \frac{2\tilde\gamma (\tilde x_0-x)}{T^*(x)} \right]  \Big ]}.
\label{19a}
		\end{eqnarray}
\\
\bigskip

{\bf B.1)  LSCO, Bi2201 and Hg1201}

\begin{figure}[tbh]
	\centering
	\includegraphics[scale=0.30]{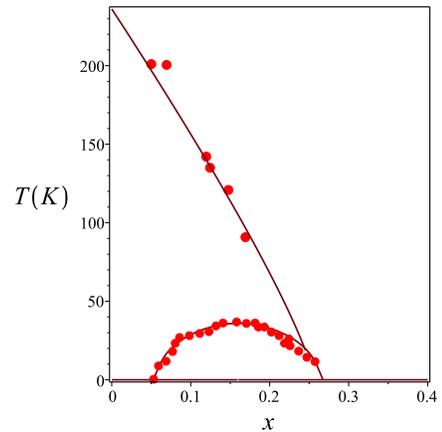}
	\caption{Solution of Eq. (\ref{16}) for the SC dome of LSCO, together with the solution of (\ref{19a}) for the pseudogap temperature $T^*(x)$. Experimental data for $T_c(x)$ from \cite{1,2,3,4} and for $T^*(x)$ from \cite{77}.} \label{figa}
\end{figure}

\begin{figure}[tbh]
	\centering
	\includegraphics[scale=0.30]{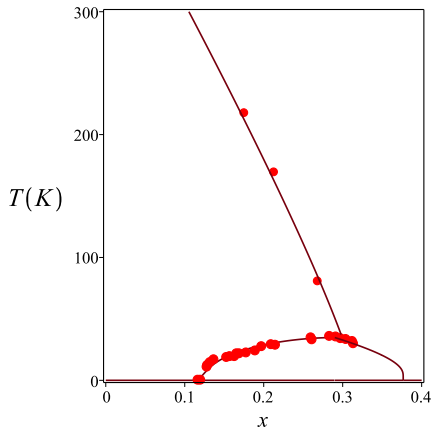}
	\caption{Solution of Eqs. (\ref{18}) and (\ref{19})  for the SC dome of Bi2201, together with the solution of (\ref{19a}) for the pseudogap temperature $T^*(x)$. Experimental data for $T_c(x)$ from \cite{1,2,3,4} and for $T^*(x)$ from \cite{7}.} \label{figb}
\end{figure}

\begin{figure}[tbh]
	\centering
	\includegraphics[scale=0.30]{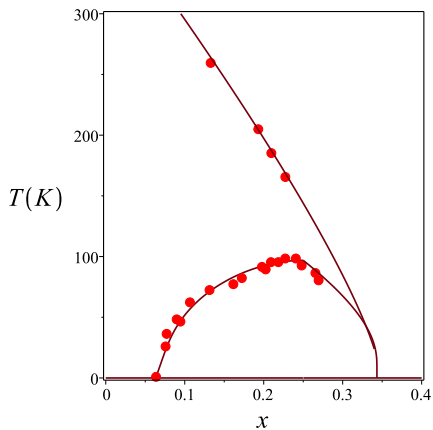}
	\caption{Solution of  Eqs. (\ref{18}) and (\ref{19})  for the SC dome of Hg1201, together with the solution of (\ref{19a}) for the pseudogap temperature $T^*(x)$. Experimental data for $T_c(x)$ and for $T^*(x)$ both from\cite{7,8}.} \label{figc}
\end{figure}

We now show in Fig. \ref{figa}, Fig. \ref{figb} and Fig. \ref{figc} the solution of (\ref{19a}) for the PG temperature,  $T^*(x)$, for  LSCO, Bi2201 and Hg1201, respectively. The SC dome is displayed in the same figure.

The PG temperature for LSCO was obtained with $\tilde\gamma=0.20\ eV$ and $\tilde x_0 = 0.255$, which implies
$\eta(g_1)= 0.01475$ and $g_1=0.31226 \ eV^{-1}$.

The PG temperature for Bi2201 was obtained with $\tilde\gamma=0.372\ eV$ and $\tilde x_0 = 0.315$, wich implies 
$\eta(g_1)= 0.00731$ and $g_1=0.31769 \ eV^{-1}$.

The PG temperature for Hg1201 was obtained with $\tilde\gamma=0.186\ eV$ and $\tilde x_0 = 0.345$, which implies
$\eta(g_1)= 0.53$ and $g_1=1.66 \ eV^{-1}$. \\
\bigskip

Notice that the ratio between the two couplings, $g$ and $g_1$ remains the same for any of the three compounds,  LSCO, Bi2201 and Hg1201: for the three of them we have $g/g_1\simeq 1.28$. This fact suggests the existence of a universal relation between the couplings responsible for the formation of the SC gap and the Pseudogap, which should originate in the underlying interaction leading to (\ref{0}).

{\bf C) Increase of $T_{max}$ with The Number of Adjacent Planes}

 It is an evident experimental fact that the optimal transition temperature becomes higher as one increases the number of   $CuO_2$ planes per primitive unit cell. Bi2201 and Hg1201, for instance, are single-layered materials, which have multi-layered relatives with a higher optimal temperature.

The mercury family, for instance, consists of \cite{7,8,mer1,mer2}: Hg1201 (single-layered) ($T_{max} = 97\ K$), Hg1212 (double-layered) ($T_{max} = 125\ K$), Hg1223 (triple-layered) ($T_{max} = 134\ K$), Hg1234 (four-layered)($T_{max} = 127\ K$) and Hg1245 (five-layered)($T_{max} = 120\ K$).  It shows an increase of the optimal temperatures as the number of adjacent layers is increased from $N=1$ to $N=3$. Then for $N=4,5$, $T_{max} $ stabilizes at a temperature approximately corresponding to $N=2$.

For the bismuth family, conversely, we have \cite{honma,honma1,bis1,bis2}
Bi2201 (single-layered) (Tmax =
34\  K), Bi2212 (double-layered) (Tmax = 92\  K), Bi2223
(triple-layered) (Tmax = 108\  K).

 From (\ref{14a}), we see that, assuming that $g$ and $g_c$ are the same for all members of a family, we may express the optimal temperature of a multi-layered cuprate with $N$ adjacent $CuO_2$ planes in terms of the same temperature of the  single-layered one, as
\begin{eqnarray}
T_{max} (N)=\frac{\eta(N)}{\eta(1)} T_{max} (1).
\label{21}
\end{eqnarray}
 Observing that $\eta(N)$ is a monotonically increasing function of $N$, the obvious effect of increasing the number of adjacent planes is to increase $T_{max}$.
This follows directly from the enhancement of the coupling parameter, namely: $g  \rightarrow N g$.

For the bismuth family: $\eta(1)=0.22666$, $\eta(2)=0.61333, \eta(3)=0.74222 $, which according to (\ref{21})  gives
\[
 T_{max} (2) = 91.99\  K,\ \ \  T_{max}^{exp} (2) = 92\ K
\]
\[
T_{max} (3)= 111.33 \  K.  \ \ \  T_{max}^{exp} (3) = 108\ K
\]
 The first values in each line above correspond to our theoretical expression (\ref{21}).
These should be compared with the experimental values \cite{honma,honma1,bis1,bis2}, appearing on the right of each line.

Accordingly, for the mercury family we have: $\eta(1)=0.63398$, $\eta(2)=0.81699, \eta(3)=0.87799, \eta(4)=0.90849, \eta(5)= 0.92679 $, which implies
\[
T_{max} (2)= 125.0\  K, \ \ \  T_{max}^{exp} (2) = 125.0\ K
\]
\[
T_{max} (3)= 134.33\  K, \ \ \  T_{max}^{exp} (3) = 134.0\ K
\]
\[
T_{max} (4)=  139.00\  K, \ \ \  T_{max}^{exp} (1) = 127.0\ K
\]
\[
T_{max} (5)= 141.80\  K, \ \ \  T_{max}^{exp} (1) = 120.0\ K
\]
The first values in each line above correspond to our theoretical expression (\ref{21}), while the 
 latter values are the experimental results  \cite{honma,honma1,mer1,mer2}.
We see that
our theoretical values for the optimal temperature of the multi-layered members of the Bi and Hg families, are in excellent agreement with the experimental values for $N=2$. Then, for $N=3$, the agreement is fairly good, whereas for $N>3$, there is no agreement.
The discrepancy, which starts to show at $N=3$ and increases for larger $N$'s can be ascribed to another effect that evidently must be taken into account as we increase the number of planes. This is the distance of such planes to the charge absorving atoms doped into the system, which becomes progressively larger as the number of planes increases. Indeed, for $N=1$ we have the two ``charge reservoir'' regions adjacent to the unique $CuO_2$ plane. For $N=2$ still each of the two planes is adjacent to a charge reservoir. Then, for $N=3$ one of the planes is no longer adjacent to any charge reservoir, while for $N=4$ and $N=5$ the innermost planes are located far away from the charge reservoirs. It happens that while the outer planes are optimally doped the inner planes are poorly doped and, consquently, remain, to a large extent, underdoped \cite{ml}. The number of active $CuO_2$ planes, namely, the ones that are adjacent to a charge reservoir, in this case, is equivalent to the one we have for $N=2$, hence the temperature stabilizes at values similar to the ones we had for $N=2$.

\bigskip

{\bf D) Superposition of the SC and AF Domes}

An evident effect in the phenomenology of multi-layered cuprates is the observation that, as we increase the number of $CuO_2$ planes per primitive unit cell, the SC and AF domes come closer to each other and eventually superimpose \cite{ml}.
From (\ref{20}), we clearly see that the quantum critical point $x^{\pm}_{SC}$ decreases as we increase $N$, the number of adjacent planes, as a result of the enhancement of the effective coupling constant $g$. Hence it will eventually come inside the AF dome that will consequently be superimposed to the SC dome.
Indeed, using (\ref{17}), (\ref{20}) and considering that $\eta(N)$ is a monotonically increasing function of $N$ we can understand why $x^{-}_{SC}(N)$ decreases as $N$ increases, therefore producing a superposition of the SC and AF domes. Furthermore, since the innermost planes are far from the charge reservoirs, they remain underdoped, and consequently in the AF phase while the outermost planes are efficiently doped and go into the SC phase.\\
\bigskip

{\bf Discussion}

 Our results indicate that the mechanism of Cooper pair formation in the cuprates must produce an  effective theory for the holes doped into the system, whose Hamiltonian is defined on the oxygen lattice of the $CuO_2$ planes. Because of the alternate overlap between $p_x$ and $p_y$ orbitals with the $Cu^{++}$ ion orbitals, which forms bridges for the doped holes, the oxygen lattice splits into two inequivalent sublattices. Cooper pairs are formed by holes belonging to different sublattices, naturally yielding a d-wave SC order parameter. Pseudogap phenomena, conversely, can be ascribed to a hole-repulsive term, favoring exciton formation, which condense in a d-wave symmetric gap (DDW) in a PG phase which competes with the Cooper pair formation of the SC phase. The d-wave character of both order parameters originates from the splitting of the oxygen lattice. For the case when $g=g_1$, this model  provides a SC pairing interaction derived from the spin-fermion model \cite{ecm1}, the two sub-lattices, $p_x$ and $p_y$ of the oxygen lattice being a realization of the two fermion species contained in that model. The experimental observation of co-existence of the SC and PG d-wave order parameters would be a clear sign of that.

By integrating over the fermions and minimizing the resulting effective action, we derive implicit equations, both for the critical SC temperature $T_c(x)$ and for the PG critical temperature $T^*(x)$, as a function of doping.   The solution of such equations is, then compared with the experimental data for different compounds, showing an excellent agreement.
 The increase of $T_{max}$ with the number of adjacent planes  as well as the superposition of SC and AF domes in multi-layered cuprates can be understood as a consequence of the enhancement of the coupling $g\rightarrow Ng$ produced by the presence of these planes. As $N$ increases, however, the inner planes progressively recede from the charge reservoirs, an effect that counteracts the enhancement o the coupling parameter, thus leading to a stabilization (or even decrease) of $T_{max}$ as we increase $N$. Based on our results one can devise a way to increase $T_c$ in cuprates: this would be achieved by effectively doping the innermost planes in multilayered cuprates. For that purpose, one should design materials with a unit cell containing as much layers as possible but with charge reservoirs intercalating no more than two layers. This would neutralize the above effect, thereby increasing $T_c$.

It would be quite interesting to investigate the possible relation of our expression for $T_{max}$, with the formula derived in \cite{R1}, on the basis of a Coulomb interaction.

Our results open a new avenue of investigation of the physical properties of High-Tc cuprates, with outstanding possibilities. Among these, how to describe the pseudogap transition and the charge ordering phases within this framework, how to include the antiferromagnetic phase in the picture, how to describe the interplay of the AF and SC phases, how to describe the resistivity above $T_c$. 
 The crucial issue in high-Tc superconductivity, of course, remains the underlying mechanism of pair formation. The results reported here could be a concrete step forward towards the complete understanding of the nature of this mechanism. \\
\bigskip

  $$
\\ 
$$

{\bf Acknowledgments}

We thank A. V. Balatsky and C. Morais Smith for stimulating conversations. We would also like to thank E. Fradkin, S. Kivelson, J. Tranquada and P. A. Marchetti, for very useful comments.
E. C. Marino was supported in part by CNPq and by FAPERJ. V. S. Alves acknowledges CNPq for support. Reginaldo de Oliveira Jr acknowledges CAPES and FAPERJ for support. \\
\bigskip
\\
\bigskip

{\bf Competing Interests}

The authors declare that they have no competing financial or non-financial interests.\\
\bigskip 

{\bf Data Availability Statement}

The authors declare that all data supporting the findings of this study are
available within the paper and its supplementary information files. The MAPLE codes leading to the curves displayed in the text are available, upon request, from the corresponding author.\\
\bigskip
\\
\bigskip

{\bf Participation of the Authors} 

E.C.M. devised and proposed the problem. E.C.M. and R.O.C.J. (under the supervision of E.C.M.) made the calculations. The Supplementary Material was prepared by R.O.C.J., L.H.C.M.N. and V.S.A. The manuscript was written by E.C.M. with input from all the authors. All the authors discussed every detail of the work.\\
\bigskip

{\it Corresponding author: ECM (marino@if.ufrj.br)}

\end{document}